\documentclass[aps,twocolumn,floats,prd,nofootinbib,superscriptaddress]
{revtex4-2}

\usepackage{amsmath,amsfonts,amssymb}
\usepackage{graphicx}

\usepackage{mathrsfs}
\usepackage{dcolumn}
\usepackage{appendix}
\usepackage{placeins}
\usepackage[usenames,dvipsnames]{xcolor} 
\usepackage[export]{adjustbox}
\usepackage{bm}
\usepackage{hyperref}
\usepackage{adjustbox}
\usepackage{tabularx}
\usepackage{subfigure}
\usepackage{multirow}
\usepackage{array}
\usepackage{float} 

\hypersetup{
	colorlinks=true,
	linkcolor=red,
	filecolor=magenta,      
	urlcolor=blue,
}
\usepackage[normalem]{ulem}

\newcommand{\RNum}[1]{\uppercase\expandafter{\romannumeral #1\relax}}

\begin{document}

\preprint{}

\title{First galaxy ultraviolet luminosity function limits on dark matter-proton scattering}
\author{Hovav Lazare}
\email{hovavl@post.bgu.ac.il}
\affiliation{Department of Physics, Ben-Gurion University of the Negev, Be’er Sheva 84105, Israel}

\author{Ely D. Kovetz}
\affiliation{Department of Physics, Ben-Gurion University of the Negev, Be’er Sheva 84105, Israel}
\affiliation{Texas Center for Cosmology and Astroparticle Physics, Weinberg Institute,
Department of Physics, The University of Texas at Austin, Austin, TX 78712, USA}

\author{Kimberly K. Boddy}
\affiliation{Texas Center for Cosmology and Astroparticle Physics, Weinberg Institute,
Department of Physics, The University of Texas at Austin, Austin, TX 78712, USA}

\author{Julian B. Mu$\Tilde{\rm n}$oz}
\affiliation{Department of Astronomy, The University of Texas at Austin, 
Austin, TX 78712, USA}
\affiliation{Cosmic Frontier Center, The University of Texas at Austin, Austin, TX 78712, USA}

\begin{abstract}

Scattering between dark matter (DM) and protons leads to suppressed small-scale fluctuations, with implications for a variety of cosmological observables. In this work, we search for evidence of  DM-proton scattering with an interaction cross section $\sigma\!=\!\sigma_0 (\frac{v}{c})^n$ for $n=0,2$ and $4$, corresponding e.g.\ to velocity-independent contact interactions from heavy mediators, velocity-dependent pseudoscalar-mediated scattering, and higher-order dipole interactions, respectively, using high-redshift ($z \sim4-10$) ultraviolet galaxy luminosity functions (UVLFs) observed by  Hubble Space Telescope (HST). We employ an adjusted implementation of \texttt{GALLUMI} combined with the modified Boltzmann solver \texttt{CLASS}  \texttt{DMeff}    that accounts for interacting DM, and incorporate UVLF data from both blank and lensed HST fields, alongside \textit{Planck} CMB data and the Pantheon supernova catalog in a Bayesian analysis framework to set constraints on $\sigma_0$. 
Our results show that including lensed UVLF data, which probe fainter galaxies than the blank HST fields and thus smaller scales, leads to a substantial improvement in the constraints on
$\sigma_0$
for $n>0$, surpassing existing bounds from Milky-Way (MW) satellite abundance and CMB anisotropies. For $m_{\chi} = 1\,\rm MeV $, for example, we set the  upper bounds at $1.1\times 10^{-25} \, \rm cm^2$ for $n=2$ and  $2.1\times 10^{-22} \, \rm cm^2$ for $n=4$. 
For $n=0$, our bound is within an order of magnitude of those from the Lyman-$\alpha$ forest and MW satellites.

\end{abstract}

\maketitle

\noindent\textit{Introduction.} Numerous cosmological and astrophysical observations have proven the need for a dark matter (DM) component in the standard model of cosmology \cite{Rees:2003tgy, Garrett:2010hd}. While the gravitational interaction of the DM fluid is well established, its additional interactions with Standard Model (SM) particles, if exist, remain unknown. A series of ground-based experiments designed to directly detect models in which massive DM particles interact weakly with SM particles~\cite{Cushman:2013zza, Battaglieri:2017aum, Akerib:2022ort} have set tight bounds on the scattering cross section for DM mass above the GeV scale, but these typically lose sensitivity at lower energies due to nuclear-recoil kinematics.

An alternative method to probe DM-nucleon scattering at lighter DM masses is to focus on cosmological and astrophysical observables that are sensitive to the suppression of small-scale matter fluctuations resulting from the exchange of heat and momentum between the interacting cosmological fluids~\cite{Chen:2002yh, Dvorkin:2013cea, Gluscevic:2017ywp, Boddy:2018kfv, Boddy:2018wzy, Boddy:2022tyt, Li:2022mdj, Slatyer:2018aqg, Buen-Abad:2021mvc, He:2025npy, Boehm:2004th, Nguyen:2021cnb, Maamari:2020aqz,  Rogers:2021byl,Xu:2018efh,  Becker:2020hzj, Nadler:2019zrb, DES:2020fxi}.

For a broad class of DM models, the momentum-transfer cross section can be parametrized as $\sigma \!= \!\sigma_0 v^n$ (throughout this work we set $c=1$), where different values of $n$ correspond to different physical scenarios~\cite{Boddy:2018kfv}.
Non-negative values of $n$ imply an early ($z \gg 10^5$) coupling between DM and SM particles, which suppresses clustering in modes that enter the horizon before the DM fluid decouples and introduces an oscillatory feature in the matter power spectrum at small scales. Bounds on  $\sigma_0$ have been derived from various observables, such as
cosmic microwave background (CMB) anisotropies \cite{Gluscevic:2017ywp, Boddy:2018kfv, Boddy:2018wzy}, 
 the Milky Way (MW) satellites \cite{Nadler:2019zrb, Maamari:2020aqz, DES:2020fxi} and the Lyman-$\alpha$ forest \cite{Xu:2018efh, Rogers:2021byl}. All  these bounds stem from the suppression of small scale matter fluctuations, each probe sensitive to a different range of scales at distinct redshifts. 
In this {\it Letter}, we study a model in which DM is allowed to scatter elastically with protons\footnote{Interactions with protons in helium nuclei are ignored here.}
 (i.e.\ interacting dark matter - IDM). This interaction is studied through its effect on ultraviolet (UV) galaxy luminosity functions (UVLFs) at high redshifts ($z=4\!-\!10$), measured by the Hubble Space Telescope (HST)~\cite{Bouwens:2021abc, Oesch:2017abc}. In addition, we examine the impact of including UVLFs from faint, lensed galaxies~\cite{Bouwens:2022ojz} on the derived limits. These galaxies are magnified by frontier fields clusters observed by HST at redshifts $z=2\!-\!9$~\cite{Bouwens:2021abc, Oesch:2017abc}.
 As the lensed UVLFs allow probing fainter galaxies residing in smaller DM halos, they extend the sensitivity to smaller scales, which can be affected by more modest DM-proton interactions.

\begin{figure*}[t]
	\centering
	\includegraphics[width = \textwidth]{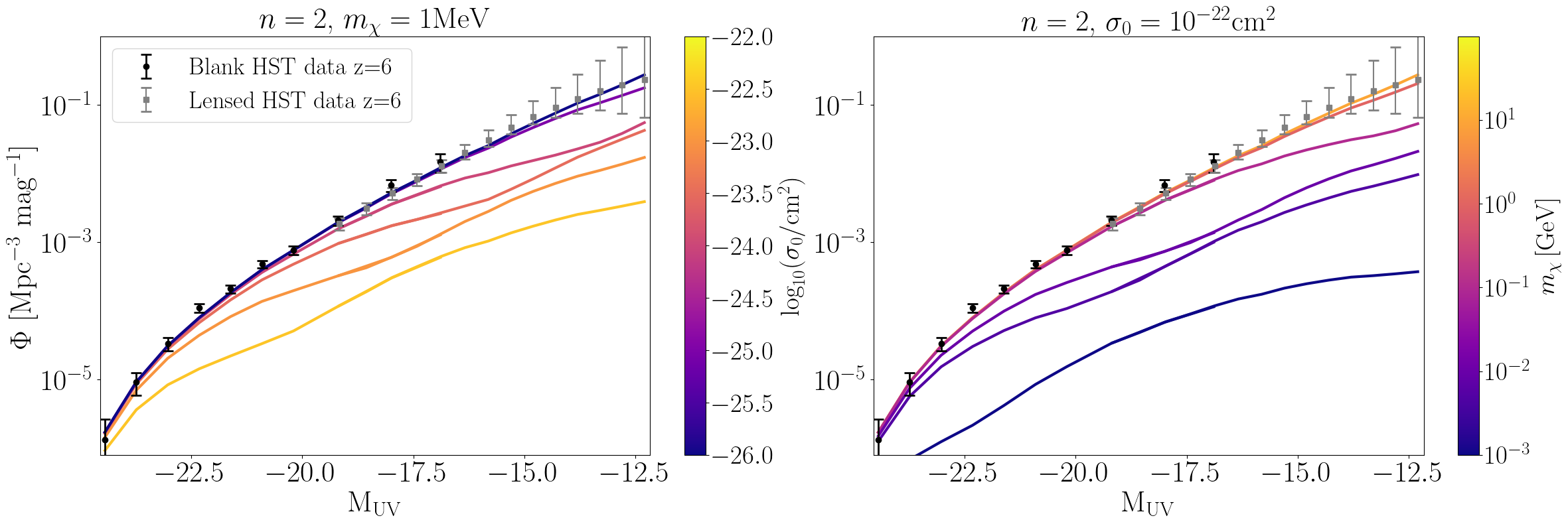}
    \vspace{-12pt}
	\caption{The impact of IDM on the UVLFs at $z=6$. {\it Left:} The scattering cross section is varied while setting $n=2$ and $m_\chi = 1 \, \mathrm{MeV}$. Increasing the interaction strength impedes structure formation at increasingly larger scales.
     {\it Right:} Here we set $n=2$ and $\sigma_0 = 10^{-22} \, \mathrm{cm^2}$ and vary $m_\chi$. Decreasing the DM particle mass has a similar effect as increasing the cross section,  both leading to a deviation from the HST measurements. The other astrophysical and cosmological parameters used to generate these figures are taken to be the best fit values from the MCMC posteriors described below. The figures show both data from the HST blank fields and the lensed fields.
     We point out that besides the noticeable effect of the reduction in the galaxies number density, the dark acoustic oscillations in the matter power spectrum also lead to small oscillatory features in the UVLFs, which are distinct from the small-scale suppression effect of other DM models (i.e. Fig. 2 in Ref.~\cite{Lazare:2024uvj} for FDM).
     }
\label{fig:LF_z=7}
\end{figure*}

Previous studies have used UVLFs to probe DM candidates that feature small scale suppression, such as warm DM (WDM) \cite{Menci:2016eui, Rudakovskyi:2021jyf, Liu:2024edl, Corasaniti:2016epp, Ellis:2025xju} and fuzzy DM (FDM)~\cite{Bozek:2014uqa, Schive:2015kza,Corasaniti:2016epp, Ellis:2025xju,
Lazare:2024uvj,  Winch:2024mrt}, as well as to set model-agnostic limits on the matter power spectrum~\cite{Sabti:2021unj}, yielding tight bounds on deviations from CDM. 
Our UVLF constraints on IDM significantly tighten current bounds on $n > 0 $ models, and complement those for the velocity-independent ($n=0$) case.

\noindent\textit{Method.}  For the modeling of the UVLFs, we use a modified version of \texttt{GALLUMI}\footnote{\href{https://github.com/NNSSA/GALLUMI_public}{github.com/NNSSA/GALLUMI\_public}} \cite{Sabti:2021unj, Sabti:2021xvh, Sabti:2023xwo}, which incorporates three main components. The first is  a modification to the halo mass function (HMF), which describes the number density of dark matter halos as a function of their mass and depends on the window function used to smooth the density field\footnote{\texttt{GALLUMI} uses the Sheth-Tormen mass function~\cite{Sheth:1999mn}, an extension of the Press-Schechter formalism~\cite{Press:1973iz} based on fits to simulations.}. Instead of the commonly used top-hat filter function in standard cosmology,
we implement a smooth-k filter function. The top-hat 
form is inadequate for a matter power spectrum that exhibits a cutoff (e.g.\ WDM, FDM), as modes below the cutoff scale do not feature a sharp enough cutoff in the HMF~\cite{Leo:2018odn, Schneider:2014rda, Benson:2012su}, resulting in an overestimate of low-mass halos compared to simulations.
Instead, for such models, the sharp-k filter, $W_{\text{SRK}}(k;R) = \Theta(1 - kR/c)$, is adopted \cite{Schneider:2014rda, Benson:2012su},
where $\Theta$ is the Heaviside function, and $c$ \cite{Schneider:2014rda} rescales the size of the collapsing region and is calibrated using N-body simulations.
In IDM cosmology with non-negative velocity power index, as in WDM and FDM models, the matter power spectrum exhibits a cutoff. On the other hand, unlike these models, it also features notable oscillations below the cutoff scale known as dark acoustic oscillations (DAOs)~\cite{Cyr-Racine:2013fsa}. Using the sharp-k filter in IDM cosmology propagates these oscillatory features on to the HMF, and subsequently to the UVLFs. This may lead to nonphysical results, as such oscillations are not observed in HMFs derived from simulations of cosmologies exhibiting DAOs \cite{Vogelsberger:2015gpr}. 
The smooth-k filter, defined as $W_{\text{SMK}}(k;R) = 1/(1+\left( kR/c \right )^\beta),$ interpolates between the top-hat and sharp-k filters~\cite{Leo:2018odn} (see also Ref.~\cite{Ellis:2025xju}).
The parameter $c$ has the same role as in the sharp-k filter, $\beta$ controls the sharpness of the filter, and in principle both should be calibrated to N-body simulations.
Note that a large enough value of $\beta$ reproduces the sharp-k filter. Ref.~\cite{Verwohlt:2024efh} have fitted the smooth-k filter to a set of N-body ETHOS simulations \cite{Vogelsberger:2015gpr}, including both $\Lambda \mathrm{CDM}$ cosmology and models that feature DAOs. They found that $c = 3.6$, $\beta =3.6$ can provide a good fit to all those simulations together, and we adopt these values here. Our results below will be calculated using the smooth-k filter, with the top-hat filter used for comparison and reference.

The second component is the mapping between halo mass and UV magnitude, captured by a  probability density function for the UV magnitude given the halo mass, $P(M_\mathrm{UV},M_\mathrm{h})$. In \texttt{GALLUMI} this is 
chosen to be a Gaussian distribution with width $\sigma_{\mathrm M_\mathrm{UV}}$~\cite{Shen:2023cva} and mean $\overline{\mathrm{M_{UV}}}(\mathrm{M_h})$, which  corresponds to a lognormal scattering in the halo mass-luminosity relation, because the UV magnitude scales logarithmically with the luminosity. A similar approach has been advocated in recent studies to study the JWST galaxy population, e.g.~\cite{Munoz:2023cup,Sun:2023ocn,Davies:2025wsa,Driskell:2024}.
 The width $\sigma_{\mathrm M_\mathrm{UV}}$ is set as a free parameter, while the mean is computed by mapping a given halo mass to a corresponding total stellar mass $M_\star$
 via a double power law,
 \vspace{-7pt}
 \begin{align}
    \label{eq:f_star}
    M_\ast = \dfrac{\epsilon_*}{\left(\dfrac{M_\mathrm{h}}{M_p}\right)^{\alpha_*}+\left(\dfrac{M_\mathrm{h}}{M_p}\right)^{\beta_*}} M_\mathrm{h}\ ,
\end{align}
where $\alpha_\ast > 0 , \beta_\ast<0$ are power law indices, $M_p$ is a pivot mass differing between mass regimes controlled by $\alpha_\ast$ and $\beta_\ast$, and $\epsilon_\ast$ is an overall amplitude. Both $M_p$ and $\epsilon_\ast$ parameterizations have redshift dependence,  
$
    \log_{10}\frac{M_p(z)}{M_\odot} =  M_p^\mathrm{i} + M_p^\mathrm{s}\times \log_{10}\left(1+z/1+6\right)$, 
$ \log_{10}\epsilon_*(z) = \epsilon_*^\mathrm{i} + \epsilon_*^\mathrm{s}\times \log_{10}\left(1+z/1+6\right)$. The scaling of $\alpha_\ast, \beta_\ast, \epsilon_\ast$ and $M_p$ with redshift, is obtained by an  analysis performed in Ref.~\cite{Sabti:2021xvh}, which varied each parameter independently for each redshift, and deduced its redshift evolution.
 The parametrization of the stellar to halo mass ratio, is based on simulations, and on observed constraints on the stellar to halo mass ratio 
 \cite{Wechsler:2018pic, Coupon:2015rua, Shuntov:2022qwu, Salucci:2018hqu, Sun:2016mnras,Behroozi:2019mnras, Moster:2018mnras}, which peaks at a certain halo mass. Suppression of star formation in small halos is due to supernovae feedback \cite{Dekel:1986ehj, Kay:2001hq} and photoionization heating from the UV background at the epoch of reionization \cite{Efstathiou:1992zz}. The suppression at the high mass end is less clear and is typically related to virial shocks \cite{Birnboim:2003xa} and AGN feedback \cite{Croton:2005hbr, Fabian:2012xr}.
 We note that this form is observed at low redshifts, and here we assume that it can be extrapolated to higher redshifts. We then compute the star formation rate (SFR) as $\dot{M_{\ast}}(M_{\rm h},z) =  M_\ast(M_\mathrm{h})/t_\ast H(z)^{-1}$ 
 where $t_\ast$ is a model parameter that accounts for the star formation timescale. Since $t_\ast$ and $\epsilon_\ast$ are fully degenerate, we vary the effective parameter $\epsilon_{\ast, \rm eff}^i := \epsilon_\ast^i - \log t_\ast$. The SFR is
  related to the galaxy UV luminosity through $\dot{M}_* = \kappa_\mathrm{UV}L_\mathrm{UV}$~\cite{Madau:1997pg,Kennicutt:1998zb}, where $\kappa_\mathrm{UV} = 1.15\times 10^{-28}\, M_\odot\mathrm{\,s\, erg}^{-1}\mathrm{yr}^{-1} $ is obtained from a stellar population synthesis, assuming a Salpeter initial mass function and a constant star formation rate. The mean value of the UV magnitude $M_{\rm UV}$ given $M_h$ is related to the UV luminosity via the AB magnitude~\cite{Oke:1983nt}.

Thirdly, when computing the UVLF $\Phi_\mathrm{UV}$ as an integral
\vspace{-15pt}
\begin{equation}
\label{eq:integrated_LF}
\begin{split}
    \hspace{0pt} \Phi_\mathrm{UV}(M_\mathrm{UV}) = \frac{1}{\Delta M_\mathrm{UV}}\int\limits_0^\infty \mathrm{d}M_\mathrm{h} \times \\
    \hspace{0pt} \left[\frac{\mathrm{d}n_\mathrm{h}}{\mathrm{d}M_\mathrm{h}}\, f_{\mathrm{duty}} (M_\mathrm{h}) \int\mathrm{d}M_\mathrm{UV}'P(M_\mathrm{UV}', M_\mathrm{h})\right] \, 
    \end{split}
\end{equation}
over the UV magnitude in bins of width $\Delta M_{\rm UV}$, we include the galaxy duty cycle, $f_{\mathrm{duty}}(M_{\rm h} ) = {\rm exp}(-\frac{M_{\rm turn}}{M_{\rm h}} )$, which accounts for inefficient galaxy formation in small halos, due to feedback processes and inefficient gas cooling~\cite{Shapiro:1993hn, Hui:1997dp, Barkana:2000fd, Springel:2002ux, Okamoto:2008sn, Mesinger:2008ze, Sobacchi:2014rua, Sobacchi:2015gpa}, parametrized as an exponential cutoff for halo masses below a characteristic scale $M_{\rm turn}$.

Finally, to study the effect of IDM on the UVLFs, we interface our modified {\tt GALLUMI} code with the \texttt{DMeff}\footnote{
\href{https://github.com/kboddy/class_public/tree/dmeff}{github.com/kboddy/class\_public/tree/dmeff}  (see Refs.~\cite{Gluscevic:2017ywp, Boddy:2018kfv})} modified version of the Boltzmann code \texttt{CLASS} \cite{Lesgourgues:2011re, Blas:2011rf}. 

\noindent\textit{Data.} The UVLF data we exploit in this study comes from the HST blank~\cite{Bouwens:2021abc, Oesch:2017abc} and lensed~\cite{Bouwens:2022ojz} fields (see also Ref.~\cite{Finkelstein:2022}). Ref.~\cite{Bouwens:2021abc} collected more than 24,000 sources to determine the UVLFs, where most of them are found at low redshifts ($z \sim 2-6$), yielding strong bounds on the number density of luminous 
galaxies at these redshifts, especially at the center of the observed brightness range.
The lensed UVLFs are composed of rare, faint galaxies, which 
are found behind the Hubble frontier field clusters, and are magnified by them.
They are obtained from only $\sim 2500$ sources, which results in a larger uncertainty on the galaxies number density. Their importance in the analysis stems from the fact that while the blank fields can access relatively bright sources ($\rm M_{UV} < -17$), the lensing of the faint sources ($\rm M_{UV} < -12$) opens a window to much smaller scales, which are sensitive to smaller DM-proton scattering cross sections than the blank fields. 
In order to avoid modeling the covariance between magnitude bins for the lensed UVLFs, we choose to use the binned UVLFs rather than the forward modeled UVLFs (i.e.~tables 4 and 3 in Ref~\cite{Bouwens:2022ojz}, respectively). We note that using the forward modeled UVLFs results in very slightly stronger constraints than the binned UVLFs (a factor of $\sim1.5-2$ across all $n$ and $m_\chi$ values).

Before utilizing the HST data, we account for three observational effects: (i) Cosmic variance, (ii) Dust attenuation and (iii) the Alcock-Paczynski effect.
Cosmic variance is address by imposing a minimal error of 20\% on the data, Dust attenuation is modeled using the IRX-$\beta$ relation, and the Alcock-Paczynski effect is accounted for by reevaluating the volume of the survey for each set of cosmological parameters; more details can be found in Ref.~\cite{Sabti:2021xvh}.

Fig.~\ref{fig:LF_z=7} shows the impact of various values of IDM parameters on the UVLFs, against HST data. 
As evident in the figure, 
at greater cross sections and lower DM masses,
the small scale suppression of IDM can have a major effect on the UVLFs.
Increasing the cross section and decreasing the DM particle mass both result in fluctuations of larger modes being washed out by scattering between baryons and DM, leading to a deficit of luminous galaxies with respect to HST measurements.
Notably, the oscillations at small scales in the matter power spectrum are translated into weaker, but noticeable oscillations in the UVLFs. This is a unique imprint of DAOs, that can potentially be detected given high-$z$ measurements.

In this luminosity range, $\Lambda \rm CDM$ is consistent with the  smallest choice for the scattering cross section, $\sigma_0=10^{-26} \rm cm^2$ (left panel), or the highest mass, $m_\chi=10\, \rm GeV$ (right panel) curves, and will give the exact same result (the DM scattering will only impact higher magnitudes).
We note that as Fig.~\ref{fig:LF_z=7} demonstrates, overlapping magnitude bins are consistent between the blank field and lensed UVLFs, which is the case for all the redshift bins used in this work.
Importantly, accounting for the lensed data allows to further discriminate between values of the cross section (or DM mass) that would be otherwise indistinguishable.
Fig.~\ref{fig:LF_z=7} shows this effect for $n=2$ for one specific redshift bin;  a similar phenomenon occurs for $n=0,4$ and at all redshifts examined in this work.

we perform an MCMC analysis based on the Metropolis-Hastings~\cite{Hastings:1970aa} algorithm using    \texttt{MontePython}~\cite{Audren:2012wb, Brinckmann:2018cvx}, combining the UVLF  HST blank and lensed data at redshifts $z=2-10$ \cite{Bouwens:2021abc} with the \textit{Planck} CMB likelihood~\cite{Planck:2018vyg}---where we use the high-$\ell$ TT, TE, and EE angular power spectra, the low-$\ell$ EE power spectrum, and the lensing power spectrum ($\phi\phi)$---and the Pantheon type-Ia supernovae likelihood \cite{Pan-STARRS1:2017jku}.

\noindent\textit{Results.}  To place a constraint on $\sigma_0$,
We fix the velocity power index $n$ and the DM particle mass $m_\chi$ and vary the astrophysical parameters $\alpha_\ast,\beta_\ast,M_p^\mathrm{i}, M_p^\mathrm{s}, \epsilon_{*, \rm eff}^\mathrm{i}, \epsilon_*^\mathrm{s}$ used in the $M_\ast(M_h)$ relation as well as $\sigma_{M_{\rm UV}}$, $M_{\rm turn}$, and the optical depth $\tau_{\rm reio}$; Table~\ref{tab:prior_table} lists the parameters and prior ranges. We do not set priors on the cosmological parameters $\omega_{cdm},\,A_s$ and $n_s$. 
We repeat the procedure for 7 different DM masses in the range $10\,\rm keV - 10\, \rm GeV$ for each $n$, and extract constraints on $\sigma_0$ as a function of $m_\chi$ at 95\% confidence level.

\begin{figure}[b]
\vspace{-0.25in}
	\centering

	\includegraphics[width = 0.95\columnwidth]{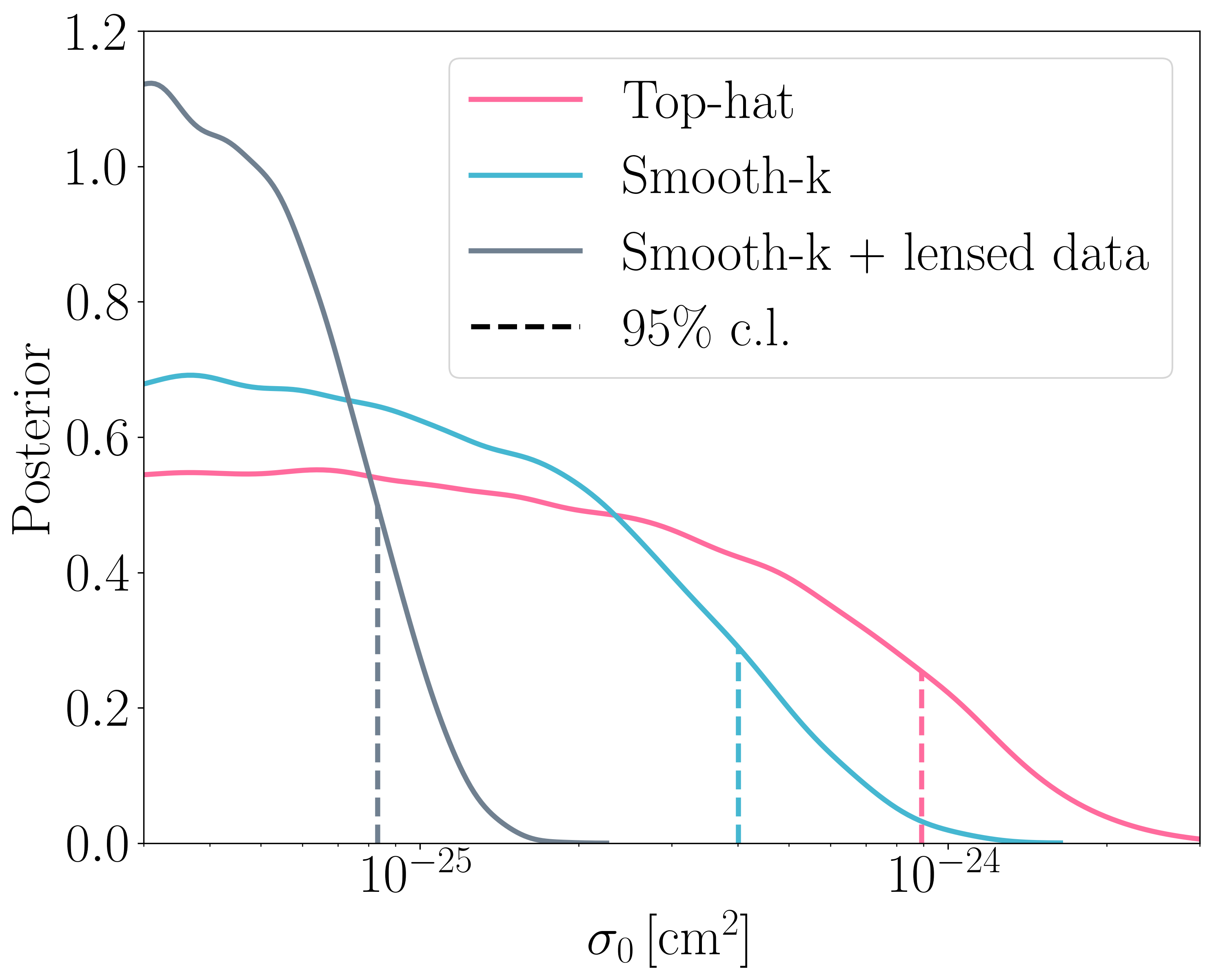}
   
	\caption{
    The $\sigma_0$ posteriors from the MCMC analysis, for $n=2$, $m_\chi = 1 \, \rm MeV$, where the  UVLFs are calculated using the top-hat filter (pink), the smooth-k filter (cyan) and the latter with both blank and lensed fields. The dashed lines represent the 95\% highest density interval, which is the value we report as the upper bound on the interaction strength.
 }
 \label{fig:constraints-TH-SK}
\end{figure}

Our 
$\sigma_0$ posteriors for $m_\chi\!=\!1\,{\rm MeV}$, are shown in Fig.~\ref{fig:constraints-TH-SK}, for the choice of a top hat and smooth-k filters. As Fig.~\ref{fig:constraints-TH-SK} implies, using the smooth-k filter improves the bounds by a factor of $\sim 2$ over the top-hat filter, which does not change our final conclusions significantly. However, since the smooth-k filter effectively interpolates between the top-hat and the sharp-k filters~\cite{Leo:2018odn}, using the latter---which is often treated as a standard choice in the context of small-scale-suppressing models---would have resulted in an even stronger bound. From this point of view, one can regard our constraints as an intermediate result between the conventional top-hat and sharp-k filter choices.

\begin{table}[h!]
\vspace{-0.05in}

\centering
    
\begin{tabular}{ |c|c|c|}
 \hline
 \multicolumn{3}{|c|}{Priors} \\
 \hline
 Parameter name &Lower bound &Upper bound\\
 \hline 

 $\tau_{reio}$ & 0.004 &  $\infty$ \\
 $\alpha_\ast$ & -3.0  &  0 \\
 $\beta_\ast$& 0  & 3.0 \\
 $\epsilon_{\ast, \rm eff}^i$& -3.0  & 3.0\\
  $\epsilon_\ast^s$& -3.0  & 3.0\\  
 $M_p^i$& 7.0  & 15.0\\
 $M_p^i$& -3.0  & 3.0\\
  $\sigma_{\mathrm{M_{UV}}}$ & 0.001  & 3.0\\

  $\log \mathrm{\frac{M_{turn}}{M_\odot}}$& 8.0  & 11.0\\
 \hline
 
\end{tabular}
\vspace{0.05in}
\caption{Prior range summary for some of the the parameters used in the MCMC runs. Prior ranges for the other parameters are mentioned in the main text.}

\label{tab:prior_table}
\vspace{-0.15in}
\end{table}

\begin{figure}[htbp]
  \centering

  \subfigure{\includegraphics[width=\linewidth]{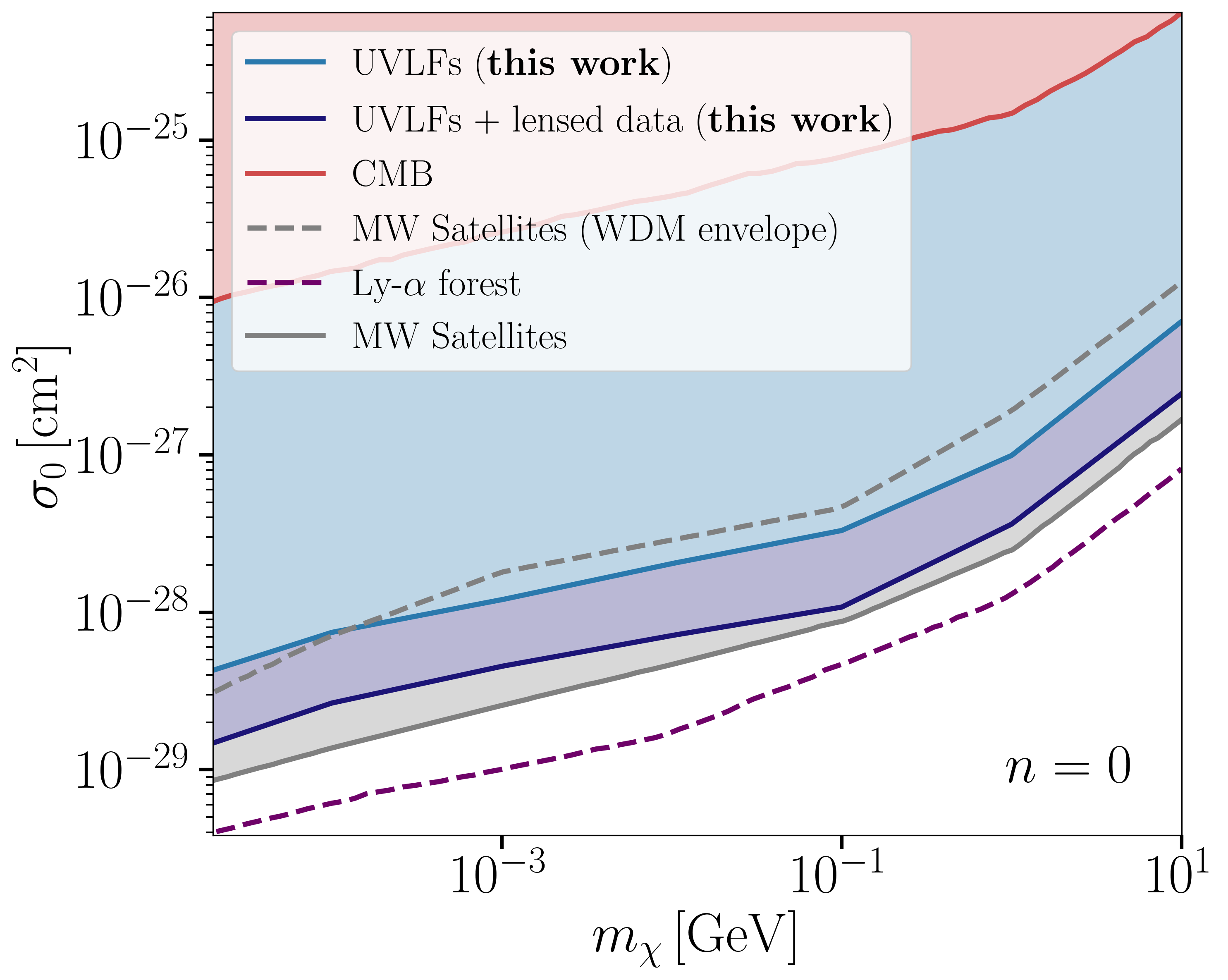}
  }
  \subfigure{\includegraphics[width=\linewidth]{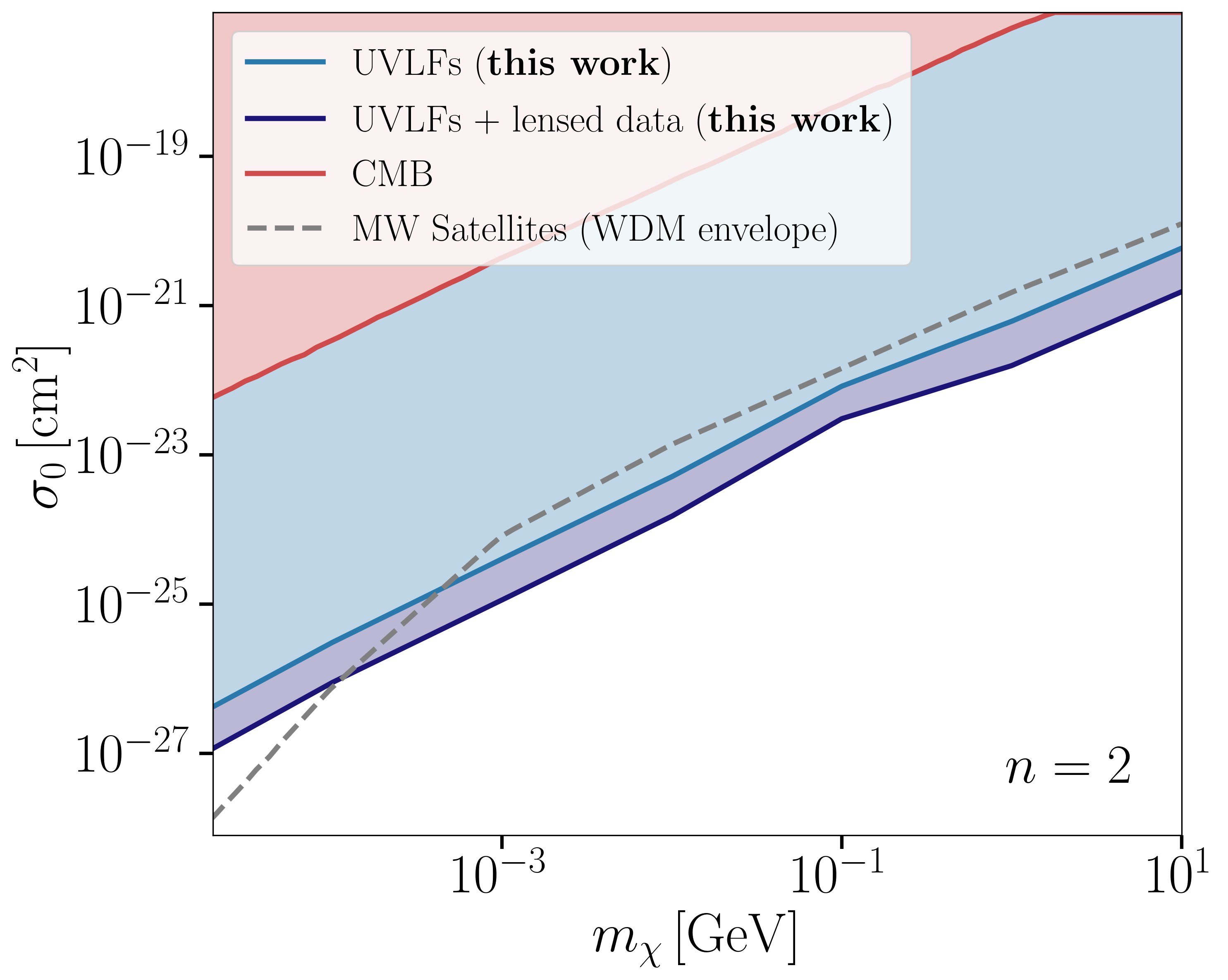} }

  \subfigure{\includegraphics[width=\linewidth]{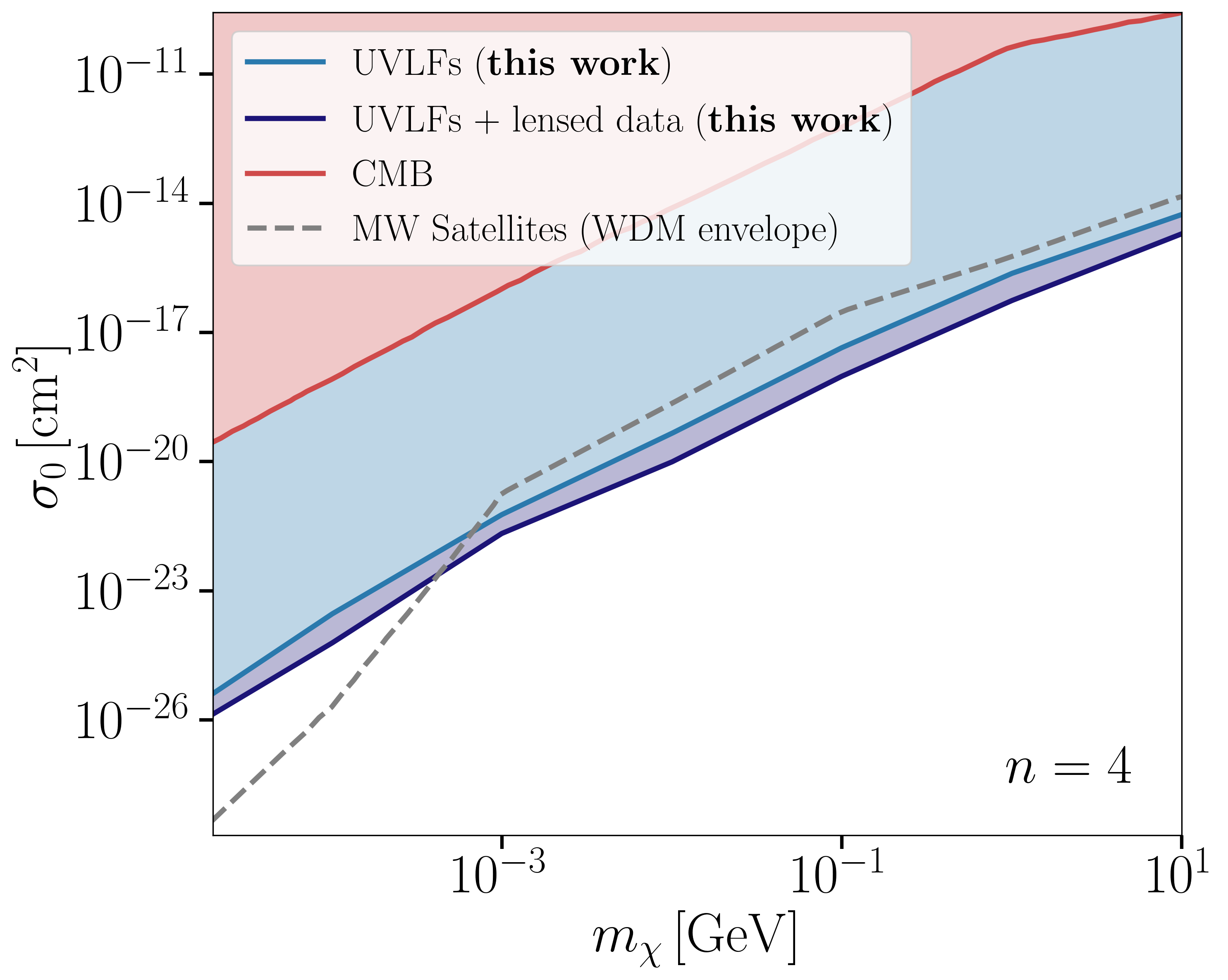}}

  \caption{Constraints on the DM-proton interaction cross section for velocity power index $n=0,2,4$ from UVLFs when using the smooth-k filter, with and without including data from the lensed field. The bounds obtained with the lensed fields show a major improvement over current constraints for $n=2,4$, and are only inferior to constraints for the Lyman-$\alpha$~\cite{Rogers:2021byl} forest for and MW satellites~\cite{DES:2020fxi} for $n=0$.}
  \label{fig:constraints-SK-lensed}
\end{figure}

Our final result, including the lensed data, is presented in Fig.~\ref{fig:constraints-SK-lensed}, where we show the bounds derived here next to constraints from other studies using various observables, for $n=0,2,4$; numerical values are provided as Supplementary Materials.
As expected, the lensed UVLFs significantly improve over the blank field bounds, achieving the strongest results to date for the $n\!=\!2$ and $4$ scenarios.  We have verified that these results do not depend on the choice of the halo-galaxy connection model, by repeating the analysis with a different model from \texttt{GALLUMI}, where the derivation of the SFR from the stellar mass is replaced with a halo accretion model (model I in \texttt{GALLUMI}). This model introduces another free parameter, and yields constrains weaker by a factor of $\sim 1.5-2$, leaving the main conclusion of this work unchanged.
We note that though the constrains we derived can potentially depend on the values of the chosen dust attenuation parameters (IRX-$\beta$ relation), in practice, these mostly affect the bright end, whereas our bounds are driven by the faint end UVLFs. This claim is also supported by a consistency test conducted in Ref.~\cite{Sabti:2021xvh}.

\noindent\textit{Discussion.}  
Surprisingly, the UVLF-dominated constraints derived here roughly match constraints that were derived from MW-satellites by Ref.~\cite{Maamari:2020aqz} (gray dashed line in Fig.~\ref{fig:constraints-SK-lensed}). This is somewhat unexpected, as MW satellites are known to probe smaller scales than UVLFs, up to $\sim 30 \rm Mpc^{-1}$. However, in the approximation implemented by Ref.~\cite{Maamari:2020aqz}, the IDM matter power spectrum was compared to that of WDM. Since the matter power spectrum in IDM cosmology and WDM cosmology take a different shape, to make such a comparison, the IDM matter power spectrum was forced to reside below the WDM one down to very small scales. In this approach, one cannot exploit the full sensitivity of the MW satellites  to small-scale suppression, and can only detect effects on scales comparable to those probed by the blank field UVLFs. 
On the other hand, Refs.~\cite{Nadler:2019zrb, DES:2020fxi} which derived limits on DM-proton scattering from MW satellites for $n=0$ (gray shaded region in the first panel of Fig.~\ref{fig:constraints-SK-lensed}), used a different fitting approach to exploit the full constraining power of MW satellites, which is not applicable for $n=2,4$.

As our UVLF constraints are  weaker than bounds from the Lyman-$\alpha$ forest 1D power spectrum~\cite{Rogers:2021byl} (for $n\!=\!0$), we conduct a consistency check to verify that UVLF data that probes small enough scales can compete with the Lyman-$\alpha$ constraints. The 1D Lyman-$\alpha$ flux power spectrum probe scales down to $\log k_f [\rm s \, km^{-1}]  \!=\! -0.7$, which corresponds to a wavenumber of $k \!\approx\! 18 \, \rm Mpc^{-1}$.\footnote{The wavenumber is approximate, as the conversion depends on the Hubble rate, which depends on cosmological parameters.}
To create mock data at smaller scales, we fit a linear function to the natural logarithm of the UVLF and its uncertainty at redshifts $z\!=\!4\!-\!6$. We use this linear function to extrapolate the UVLFs at those redshifts, extending them up to $\rm M_{UV} \!\approx\! -11$, with a spacing of $1$ mag, starting at the faintest real data point. The chosen value of $-11$ magnitudes roughly corresponds to the same wavenumber the Lyman-$\alpha$ data probes, so we expect to obtain constraints similar to those derived in Ref.~\cite{Rogers:2021byl}.
JWST will probably not be able to observe such faint galaxies directly, as current estimations suggest that it can reach $\rm M_{UV} \!\gtrsim\! -15$~\cite{Williams:2024}. However, lensed UVLFs found in HST data, already extend up to $\rm M_{UV} \!\approx\! -12$, implying that lensed UVLFs from JWST are likely to be able to probe galaxies with $\rm M_{UV} \approx -11$. 

We find that for $n\!=\!0$ and  $m_\chi \!= \!1 \, \rm MeV$, the upper bound on the interaction cross section is $\sigma_0\! <\! 1.14 \times 10 ^{-29} \, \rm cm^2$, which is equivalent to the values reported in Ref.~\cite{Rogers:2021byl}: $\sigma_0 \!<\! 1.10 \times 10 ^{-29} \, \rm cm^2$. Our constraints for $n\!=\!2$ and $4$ are the best to date. This further supports the capability of the UVLFs to constrain smaller scales with future observations of fainter galaxies by JWST.
In addition, we emphasize that an analysis pipeline based on UVLFs is computationally cheap and relies on a straightforward approach. On the other hand, deriving constraints from observables such as MW satellites and Lyman-$\alpha$ forest necessitates heavy simulations and/or approximated frameworks, such as mapping constraints from other models~\cite{Nadler:2019zrb, Maamari:2020aqz,DES:2020fxi} and Gaussian process (GP) \cite{Rasmussen:2006} based emulators trained on small datasets \cite{Rogers:2021byl}.

We note that in this analysis we did not take into account the possible impact of DM-baryon scattering on the astrophysical processes of galaxy formation. In principle, such interactions can change the accretion rates and alter the thermal history of the gas. These effects can influence the formation of galaxies and lead to a notable signature on the UVLFs. A more complete study incorporating all relevant astrophysics is deferred to future work.

Our work reflects the dawn of cosmic dawn cosmology. Looking ahead, future UVLF observations from JWST~\cite{Stark:2025aaa, Donnan:2024, Harikane:2023}, reaching fainter galaxies at higher redshifts, will offer the potential to significantly improve these constraints because of the increased sensitivity to early structure formation and the impact of IDM on smaller scales. We do not use the JWST UVLFs in this work, as the number of galaxies used to construct the high-redshift JWST UVLFs is still substantially smaller than in the HST UVLFs, making them less well established and therefore less constraining (see e.g.~\cite{Winch:2024mrt}). In addition, as indicated in recent studies (e.g.~\cite{Yung:2023bng}), the extremely high redshifts ($z \sim 14$) probed by JWST may not be adequately described by the Sheth–Tormen (ST) mass function.
This complicates the construction of a unified and reliable model applicable to both HST and JWST data, which we leave for future work.
We note that Ref.~\cite{Yung:2023bng} mentions that the ST formalism might already underpredict the HMF at $z\sim 10$ by a factor for $0.2-0.3$ dex. Such a systematic error will likely not impede the final results of our work, as the constraints are governed by the lower redshift bins, where the number of detected galaxies is much larger, making its uncertainties much smaller.

Further down the road, the 21cm signal, being sought by experiments such as HERA~\cite{DeBoer:2016tnn}, LOFAR~\cite{LOFAR:2013jil} and eventually the SKAO~\cite{Mellema:2012ht}, holds exceptional promise for probing even smaller scales than accessible with galaxy surveys, due to its sensitivity to the faintest galaxies~\cite{Flitter:2023mjj, Rahimieh:2025fsb, Rahimieh:2025lbf, Munoz:2019hjh}. The combination of these space-borne and ground-based cosmic dawn observations will provide much stronger bounds on IDM cosmology as well as other models that suppress small scale structure.

\vspace{-10pt}
\begin{acknowledgments}
\vspace{-7pt}
The authors thank Jordan Flitter, Sarah Libanore, and Subhajit Ghosh for
useful and enlightening discussions. HL
is supported by the Zin fellowship awarded by the BGU
Kreitmann School.
EDK acknowledges
 support from the U.S.-Israel Bi-national Science
Foundation (NSF-BSF grant 2022743 and BSF grant 2024193) and the Israel National Science Foundation (ISF grant 3135/25), as
well as support from the joint Israel-China   program (ISF-NSFC grant  3156/23). 
KB acknowledges support from the NSF under Grant No. PHY-2413016.
JBM acknowledges support
from NSF Grants AST-2307354 and AST-2408637, and
the CosmicAI institute AST-2421782.
HL and EDK thank the Weinberg Institute at the University of Texas at Austin for warm hospitality during student and sabbatical visits, respectively.

\end{acknowledgments}

\begin{table}[H]

\centering

\label{tab:combined_constraints}
\begin{tabular}{|c|c|c|c|c|}
\hline
\textbf{n} & $m_\chi$ & \textbf{TH [cm$^2$]} & \textbf{SMK [cm$^2$]} & \textbf{+ lensed [cm$^2$]} \\
\hline
\multirow{7}{*}{0} & 10 KeV & $7.49 \times 10^{-29}$ & $3.37 \times 10^{-29}$ & $1.15 \times 10^{-29}$ \\
 & 100 KeV & $1.19 \times 10^{-28}$ & $7.45 \times 10^{-29}$ & $2.65 \times 10^{-29}$ \\
 & 1 MeV & $2.66 \times 10^{-28}$ & $1.20 \times 10^{-28}$ & $4.53 \times 10^{-29}$ \\
 & 10 MeV & $3.31 \times 10^{-28}$ & $2.04 \times 10^{-28}$ & $7.11 \times 10^{-29}$ \\
 & 100 MeV & $5.29 \times 10^{-28}$ & $3.31 \times 10^{-28}$ & $1.08 \times 10^{-28}$ \\
 & 1 GeV & $1.70 \times 10^{-27}$ & $9.91 \times 10^{-28}$ & $3.63 \times 10^{-28}$ \\
 & 10 GeV & $1.22 \times 10^{-26}$ & $7.04 \times 10^{-27}$ & $2.45 \times 10^{-27}$ \\
\hline
\multirow{7}{*}{2} & 10 KeV & $4.78 \times 10^{-27}$ & $1.81 \times 10^{-27}$ & $4.89 \times 10^{-28}$ \\
 & 100 KeV & $6.75 \times 10^{-26}$ & $3.05 \times 10^{-26}$ & $8.87 \times 10^{-27}$ \\
 & 1 MeV & $8.95 \times 10^{-25}$ & $4.03 \times 10^{-25}$ & $1.14 \times 10^{-25}$ \\
 & 10 MeV & $1.15 \times 10^{-23}$ & $5.09 \times 10^{-24}$ & $1.51 \times 10^{-24}$ \\
 & 100 MeV & $1.89 \times 10^{-22}$ & $8.30 \times 10^{-23}$ & $3.05 \times 10^{-23}$ \\
 & 1 GeV & $1.54 \times 10^{-21}$ & $6.15 \times 10^{-22}$ & $1.57 \times 10^{-22}$ \\
 & 10 GeV & $1.87 \times 10^{-20}$ & $5.87 \times 10^{-21}$ & $1.53 \times 10^{-21}$ \\
\hline
\multirow{7}{*}{4} & 10 KeV & $3.15 \times 10^{-26}$ & $6.58 \times 10^{-27}$ & $2.66 \times 10^{-27}$ \\
 & 100 KeV & $9.26 \times 10^{-24}$ & $2.87 \times 10^{-24}$ & $6.06 \times 10^{-25}$ \\
 & 1 MeV & $1.76 \times 10^{-21}$ & $5.85 \times 10^{-22}$ & $2.14 \times 10^{-22}$ \\
 & 10 MeV & $1.38 \times 10^{-19}$ & $4.51 \times 10^{-20}$ & $9.77 \times 10^{-21}$ \\
 & 100 MeV & $1.37 \times 10^{-17}$ & $4.41 \times 10^{-18}$ & $9.56 \times 10^{-19}$ \\
 & 1 GeV & $7.40 \times 10^{-16}$ & $2.34 \times 10^{-16}$ & $5.47 \times 10^{-17}$ \\
 & 10 GeV & $1.54 \times 10^{-14}$ & $5.47 \times 10^{-15}$ & $1.95 \times 10^{-15}$ \\

\hline

\end{tabular}
\vspace{0.05in}
\caption{Constraints on the DM---proton interaction cross section derived from UV luminosity functions for different velocity power indices and DM masses. Results are shown using the top-hat and smooth-k filter, the latter also for the inclusion of lensed data.}
\end{table}


\clearpage



\begin{thebibliography}{}

\bibitem{Rees:2003tgy}
M.~J.~Rees,
``Dark matter: Introduction,''
Phil. Trans. Roy. Soc. Lond. A \textbf{361} (2003), 2427-2434
[arXiv:astro-ph/0402045 [astro-ph]].

\bibitem{Garrett:2010hd}
K.~Garrett and G.~Duda,
``Dark Matter: A Primer,''
Adv. Astron. \textbf{2011} (2011), 968283
[arXiv:1006.2483 [hep-ph]].


\bibitem{Cushman:2013zza}
P.~Cushman, C.~Galbiati, D.~N.~McKinsey, H.~Robertson, T.~M.~P.~Tait, D.~Bauer, A.~Borgland, B.~Cabrera, F.~Calaprice and J.~Cooley, \textit{et al.}
``Working Group Report: WIMP Dark Matter Direct Detection,''
[arXiv:1310.8327 [hep-ex]].

\bibitem{Battaglieri:2017aum}
M.~Battaglieri, A.~Belloni, A.~Chou, P.~Cushman, B.~Echenard, R.~Essig, J.~Estrada, J.~L.~Feng, B.~Flaugher and P.~J.~Fox, \textit{et al.}
``US Cosmic Visions: New Ideas in Dark Matter 2017: Community Report,''
[arXiv:1707.04591 [hep-ph]].

\bibitem{Akerib:2022ort}
D.~S.~Akerib, P.~B.~Cushman, C.~E.~Dahl, R.~Ebadi, A.~Fan, R.~J.~Gaitskell, C.~Galbiati, G.~K.~Giovanetti, G.~B.~Gelmini and L.~Grandi, \textit{et al.}
``Snowmass2021 Cosmic Frontier Dark Matter Direct Detection to the Neutrino Fog,''
[arXiv:2203.08084 [hep-ex]].

\bibitem{Chen:2002yh}
X.~l.~Chen, S.~Hannestad and R.~J.~Scherrer,
``Cosmic microwave background and large scale structure limits on the interaction between dark matter and baryons,''
Phys. Rev. D \textbf{65}, 123515 (2002)
[arXiv:astro-ph/0202496 [astro-ph]].

\bibitem{Boddy:2022tyt}
K.~K.~Boddy, G.~Krnjaic and S.~Moltner,
Phys. Rev. D \textbf{106}, no.4, 043510 (2022)
doi:10.1103/PhysRevD.106.043510
[arXiv:2204.04225 [astro-ph.CO]].

\bibitem{Dvorkin:2013cea}
C.~Dvorkin, K.~Blum and M.~Kamionkowski,
``Constraining Dark Matter-Baryon Scattering with Linear Cosmology,''
Phys. Rev. D \textbf{89} (2014) no.2, 023519
[arXiv:1311.2937 [astro-ph.CO]].


\bibitem{Gluscevic:2017ywp}
V.~Gluscevic and K.~K.~Boddy,
``Constraints on Scattering of keV\textendash{}TeV Dark Matter with Protons in the Early Universe,''
Phys. Rev. Lett. \textbf{121} (2018) no.8, 081301
[arXiv:1712.07133 [astro-ph.CO]].


\bibitem{Boddy:2018kfv}
K.~K.~Boddy and V.~Gluscevic,
``First Cosmological Constraint on the Effective Theory of Dark Matter-Proton Interactions,''
Phys. Rev. D \textbf{98} (2018) no.8, 083510
[arXiv:1801.08609 [astro-ph.CO]].



\bibitem{Boddy:2018wzy}
K.~K.~Boddy, V.~Gluscevic, V.~Poulin, E.~D.~Kovetz, M.~Kamionkowski and R.~Barkana,
``Critical assessment of CMB limits on dark matter-baryon scattering: New treatment of the relative bulk velocity,''
Phys. Rev. D \textbf{98} (2018) no.12, 123506
[arXiv:1808.00001 [astro-ph.CO]].


\bibitem{Li:2022mdj}
Z.~Li, R.~An, V.~Gluscevic, K.~K.~Boddy, J.~R.~Bond, E.~Calabrese, J.~Dunkley, P.~A.~Gallardo, Y.~Guan and A.~Hincks, \textit{et al.}
``The Atacama Cosmology Telescope: limits on dark matter-baryon interactions from DR4 power spectra,''
JCAP \textbf{02} (2023), 046
[arXiv:2208.08985 [astro-ph.CO]].


\bibitem{Slatyer:2018aqg}
T.~R.~Slatyer and C.~L.~Wu,
``Early-Universe constraints on dark matter-baryon scattering and their implications for a global 21 cm signal,''
Phys. Rev. D \textbf{98} (2018) no.2, 023013
[arXiv:1803.09734 [astro-ph.CO]].


\bibitem{Buen-Abad:2021mvc}
M.~A.~Buen-Abad, R.~Essig, D.~McKeen and Y.~M.~Zhong,
``Cosmological constraints on dark matter interactions with ordinary matter,''
Phys. Rept. \textbf{961} (2022), 1-35
[arXiv:2107.12377 [astro-ph.CO]].


\bibitem{He:2025npy}
A.~He, M.~M.~Ivanov, R.~An, T.~Driskell and V.~Gluscevic,
JCAP \textbf{05} (2025), 087
[arXiv:2502.02636 [astro-ph.CO]].





\bibitem{Boehm:2004th}
C.~Boehm and R.~Schaeffer,
``Constraints on dark matter interactions from structure formation: Damping lengths,''
Astron. Astrophys. \textbf{438} (2005), 419-442
[arXiv:astro-ph/0410591 [astro-ph]].

\bibitem{Nguyen:2021cnb}
D.~V.~Nguyen, D.~Sarnaaik, K.~K.~Boddy, E.~O.~Nadler and V.~Gluscevic,
``Observational constraints on dark matter scattering with electrons,''
Phys. Rev. D \textbf{104} (2021) no.10, 103521
[arXiv:2107.12380 [astro-ph.CO]].


\bibitem{Maamari:2020aqz}
K.~Maamari, V.~Gluscevic, K.~K.~Boddy, E.~O.~Nadler and R.~H.~Wechsler,
``Bounds on velocity-dependent dark matter-proton scattering from Milky Way satellite abundance,''
Astrophys. J. Lett. \textbf{907} (2021) no.2, L46
[arXiv:2010.02936 [astro-ph.CO]].






\bibitem{Nadler:2019zrb}
E.~O.~Nadler, V.~Gluscevic, K.~K.~Boddy and R.~H.~Wechsler,
``Constraints on Dark Matter Microphysics from the Milky Way Satellite Population,''
Astrophys. J. Lett. \textbf{878}, no.2, 32 (2019)
[erratum: Astrophys. J. Lett. \textbf{897}, no.2, L46 (2020); erratum: Astrophys. J. \textbf{897}, no.2, L46 (2020)]
[arXiv:1904.10000 [astro-ph.CO]].

\bibitem{DES:2020fxi}
E.~O.~Nadler \textit{et al.} [DES],
Phys. Rev. Lett. \textbf{126} (2021), 091101
[arXiv:2008.00022 [astro-ph.CO]].



\bibitem{Xu:2018efh}
W.~L.~Xu, C.~Dvorkin and A.~Chael,
``Probing sub-GeV Dark Matter-Baryon Scattering with Cosmological Observables,''
Phys. Rev. D \textbf{97}, no.10, 103530 (2018)
[arXiv:1802.06788 [astro-ph.CO]].

\bibitem{Rogers:2021byl}
K.~K.~Rogers, C.~Dvorkin and H.~V.~Peiris,
``Limits on the Light Dark Matter\textendash{}Proton Cross Section from Cosmic Large-Scale Structure,''
Phys. Rev. Lett. \textbf{128} (2022) no.17, 171301
[arXiv:2111.10386 [astro-ph.CO]].

\bibitem{Becker:2020hzj}
N.~Becker, D.~C.~Hooper, F.~Kahlhoefer, J.~Lesgourgues and N.~Sch\"oneberg,
``Cosmological constraints on multi-interacting dark matter,''
JCAP \textbf{02} (2021), 019
[arXiv:2010.04074 [astro-ph.CO]].



\bibitem{Bouwens:2021abc}
Bouwens R.~J., Oesch P.~A., Stefanon M., Illingworth G., Labb{\'e} I., Reddy N., Atek H., et al., 2021, AJ, 162, 47.

\bibitem{Oesch:2017abc}
P.~A.~Oesch, R.~J.~Bouwens, G.~D.~Illingworth, I.~Labbe, M.~Stefanon
``The Dearth of z ~ 10 Galaxies in All HST Legacy Fields—The Rapid Evolution of the Galaxy Population in the First 500 Myr",
Astrophys. J. \textbf{855}, 105 (2017) 
[arXiv:1710.11131 [astro-ph.CO]].


\bibitem{Bouwens:2022ojz}
R.~J.~Bouwens, G.~D.~Illingworth, R.~S.~Ellis, P.~A.~Oesch and M.~Stefanon,
``z {\ensuremath{\sim}} 2{\textendash}9 Galaxies Magnified by the Hubble Frontier Field Clusters. II. Luminosity Functions and Constraints on a Faint-end Turnover,''
Astrophys. J. \textbf{940}, no.1, 55 (2022)
[arXiv:2205.11526 [astro-ph.GA]].

\bibitem{Menci:2016eui}
N.~Menci, A.~Grazian, M.~Castellano and N.~G.~Sanchez,
``A Stringent Limit on the Warm Dark Matter Particle Masses from the Abundance of z=6 Galaxies in the Hubble Frontier Fields,''
Astrophys. J. Lett. \textbf{825}, no.1, L1 (2016)
[arXiv:1606.02530 [astro-ph.CO]].


\bibitem{Rudakovskyi:2021jyf}
A.~Rudakovskyi, A.~Mesinger, D.~Savchenko and N.~Gillet,
``Constraints on warm dark matter from UV luminosity functions of high-z galaxies with Bayesian model comparison,''
Mon. Not. Roy. Astron. Soc. \textbf{507} (2021) no.2, 3046-3056
[arXiv:2104.04481 [astro-ph.CO]].

\bibitem{Liu:2024edl}
B.~Liu, H.~Shan and J.~Zhang,
``New Galaxy UV Luminosity Constraints on Warm Dark Matter from JWST,''
Astrophys. J. \textbf{968} (2024) no.2, 79
[arXiv:2404.13596 [astro-ph.CO]].

\bibitem{Corasaniti:2016epp}
P.~S.~Corasaniti, S.~Agarwal, D.~J.~E.~Marsh and S.~Das,
``Constraints on dark matter scenarios from measurements of the galaxy luminosity function at high redshifts,''
Phys. Rev. D \textbf{95}, no.8, 083512 (2017)
[arXiv:1611.05892 [astro-ph.CO]].

\bibitem{Ellis:2025xju}
J.~Ellis, M.~Fairbairn, J.~Urrutia and V.~Vaskonen,
``Starlight from JWST: Implications for star formation and dark matter models,''
[arXiv:2504.20043 [astro-ph.CO]].

\bibitem{Bozek:2014uqa}
B.~Bozek, D.~J.~E.~Marsh, J.~Silk and R.~F.~G.~Wyse,
``Galaxy UV-luminosity function and reionization constraints on axion dark matter,''
Mon. Not. Roy. Astron. Soc. \textbf{450}, no.1, 209-222 (2015)
[arXiv:1409.3544 [astro-ph.CO]].

\bibitem{Schive:2015kza}
H.~Y.~Schive, T.~Chiueh, T.~Broadhurst and K.~W.~Huang,
``Contrasting Galaxy Formation from Quantum Wave Dark Matter, $\psi$DM, with $\Lambda$CDM, using Planck and Hubble Data,''
Astrophys. J. \textbf{818}, no.1, 89 (2016)
[arXiv:1508.04621 [astro-ph.GA]].



\bibitem{Lazare:2024uvj}
H.~Lazare, J.~Flitter and E.~D.~Kovetz,
Phys. Rev. D \textbf{110} (2024) no.12, 123532
[arXiv:2407.19549 [astro-ph.CO]].


\bibitem{Winch:2024mrt}
H.~Winch, K.~K.~Rogers, R.~Hlo\v{z}ek and D.~J.~E.~Marsh,
``High-redshift, Small-scale Tests of Ultralight Axion Dark Matter Using Hubble and Webb Galaxy UV Luminosities,''
Astrophys. J. \textbf{976} (2024) no.1, 40
[arXiv:2404.11071 [astro-ph.CO]].

\bibitem{Sabti:2021unj}
N.~Sabti, J.~B.~Mu{\~n}oz and D.~Blas,
``New Roads to the Small-scale Universe: Measurements of the Clustering of Matter with the High-redshift UV Galaxy Luminosity Function,''
Astrophys. J. Lett. \textbf{928}, no.2, L20 (2022)
[arXiv:2110.13161 [astro-ph.CO]].

\bibitem{Sabti:2021xvh}
N.~Sabti, J.~B.~Mu\~noz and D.~Blas,
``Galaxy luminosity function pipeline for cosmology and astrophysics,''
Phys. Rev. D \textbf{105} (2022) no.4, 043518
[arXiv:2110.13168 [astro-ph.CO]].

\bibitem{Sabti:2023xwo}
N.~Sabti, J.~B.~Mu\~noz and M.~Kamionkowski,
``Insights from HST into Ultramassive Galaxies and Early-Universe Cosmology,''
Phys. Rev. Lett. \textbf{132} (2024) no.6, 061002
[arXiv:2305.07049 [astro-ph.CO]].

\bibitem{Sheth:1999mn}
R.~K.~Sheth and G.~Tormen,
``Large scale bias and the peak background split,''
Mon. Not. Roy. Astron. Soc. \textbf{308}, 119 (1999)
[arXiv:astro-ph/9901122 [astro-ph]].


 
\bibitem{Press:1973iz}
W.~H.~Press and P.~Schechter,
``Formation of galaxies and clusters of galaxies by selfsimilar gravitational condensation,''
Astrophys. J. \textbf{187} (1974), 425-438














\bibitem{Leo:2018odn}
M.~Leo, C.~M.~Baugh, B.~Li and S.~Pascoli,
``A new smooth-$k$ space filter approach to calculate halo abundances,''
JCAP \textbf{04} (2018), 010
[arXiv:1801.02547 [astro-ph.CO]].

\bibitem{Schneider:2014rda}
A.~Schneider,
``Structure formation with suppressed small-scale perturbations,''
Mon. Not. Roy. Astron. Soc. \textbf{451} (2015) no.3, 3117-3130
[arXiv:1412.2133 [astro-ph.CO]].

\bibitem{Benson:2012su}
A.~J.~Benson, A.~Farahi, S.~Cole, L.~A.~Moustakas, A.~Jenkins, M.~Lovell, R.~Kennedy, J.~Helly and C.~Frenk,
``Dark Matter Halo Merger Histories Beyond Cold Dark Matter: I - Methods and Application to Warm Dark Matter,''
Mon. Not. Roy. Astron. Soc. \textbf{428} (2013), 1774
[arXiv:1209.3018 [astro-ph.CO]].

\bibitem{Cyr-Racine:2013fsa}
F.~Y.~Cyr-Racine, R.~de Putter, A.~Raccanelli and K.~Sigurdson,
``Constraints on Large-Scale Dark Acoustic Oscillations from Cosmology,''
Phys. Rev. D \textbf{89}, no.6, 063517 (2014)
[arXiv:1310.3278 [astro-ph.CO]].

\bibitem{Vogelsberger:2015gpr}
M.~Vogelsberger, J.~Zavala, F.~Y.~Cyr-Racine, C.~Pfrommer, T.~Bringmann and K.~Sigurdson,
``ETHOS \textendash{} an effective theory of structure formation: dark matter physics as a possible explanation of the small-scale CDM problems,''
Mon. Not. Roy. Astron. Soc. \textbf{460} (2016) no.2, 1399-1416
[arXiv:1512.05349 [astro-ph.CO]].

\bibitem{Verwohlt:2024efh}
J.~Verwohlt, C.~A.~Mason, J.~B.~Mu\~noz, F.~Y.~Cyr-Racine, M.~Vogelsberger and J.~Zavala,
``Separating dark acoustic oscillations from astrophysics at cosmic dawn,''
Phys. Rev. D \textbf{110} (2024) no.10, 103533
[arXiv:2404.17640 [astro-ph.CO]].


\bibitem{Shen:2023cva}
X.~Shen, M.~Vogelsberger, M.~Boylan-Kolchin, S.~Tacchella and R.~Kannan,
[arXiv:2305.05679 [astro-ph.GA]].



\bibitem{Munoz:2023cup}
J.~B.~Mu{\~n}oz, J.~Mirocha, S.~Furlanetto and N.~Sabti,
``Breaking degeneracies in the first galaxies with clustering,''
Mon. Not. Roy. Astron. Soc. \textbf{526}, no.1, L47-L55 (2023)
[arXiv:2306.09403 [astro-ph.CO]].

\bibitem{Sun:2023ocn}
G.~Sun, C.~A.~Faucher-Gigu{\`e}re, C.~C.~Hayward, X.~Shen, A.~Wetzel and R.~K.~Cochrane,
``Bursty Star Formation Naturally Explains the Abundance of Bright Galaxies at Cosmic Dawn,''
Astrophys. J. Lett. \textbf{955}, no.2, L35 (2023)
[arXiv:2307.15305 [astro-ph.GA]].

\bibitem{Davies:2025wsa}
J.~E.~Davies, A.~Mesinger and S.~Murray,
``Efficient simulation of discrete galaxy populations and associated radiation fields during the first billion years,''
[arXiv:2504.17254 [astro-ph.CO]].

\bibitem{Driskell:2024}
T.~Driskell, E.~O.~Nadler, A.~Benson and V.~Gluscevic,
``Population synthesis and astrophysical inference for high-$z$ JWST galaxies,''
arXiv:2410.11680 [astro-ph.GA].

\bibitem{Wechsler:2018pic}
R.~H.~Wechsler and J.~L.~Tinker,
``The Connection between Galaxies and their Dark Matter Halos,''
Ann. Rev. Astron. Astrophys. \textbf{56}, 435-487 (2018)
doi:10.1146/annurev-astro-081817-051756
[arXiv:1804.03097 [astro-ph.GA]].

\bibitem{Moster:2018mnras}
B.~P.~Moster, T.~Naab and S.~D.~M.~White,
``EMERGE - an empirical model for the formation of galaxies since z $\sim$ 10,''
Mon. Not. Roy. Astron. Soc. \textbf{477}, no.2, 1822-1852 (2018)
doi:10.1093/mnras/sty655
[arXiv:1705.05373 [astro-ph.GA]].

\bibitem{Behroozi:2019mnras}
P.~Behroozi, R.~H.~Wechsler, A.~P.~Hearin and C.~Conroy,
``UNIVERSEMACHINE: The correlation between galaxy growth and dark matter halo assembly from z = 0-10,''
Mon. Not. Roy. Astron. Soc. \textbf{488}, no.3, 3143-3194 (2019)
doi:10.1093/mnras/stz1182
[arXiv:1806.07893 [astro-ph.GA]].

\bibitem{Sun:2016mnras}
G.~Sun and S.~R.~Furlanetto,
``Constraints on the star formation efficiency of galaxies during the epoch of reionization,''
Mon. Not. Roy. Astron. Soc. \textbf{460}, no.1, 417-433 (2016)
doi:10.1093/mnras/stw980
[arXiv:1512.06219 [astro-ph.GA]].

\bibitem{Coupon:2015rua}
J.~Coupon, S.~Arnouts, L.~van Waerbeke, T.~Moutard, O.~Ilbert, E.~van Uitert, T.~Erben, B.~Garilli, L.~Guzzo and C.~Heymans, \textit{et al.}
``The galaxy{\textendash}halo connection from a joint lensing, clustering and abundance analysis in the CFHTLenS/VIPERS field,''
Mon. Not. Roy. Astron. Soc. \textbf{449}, no.2, 1352-1379 (2015)
[arXiv:1502.02867 [astro-ph.CO]].


\bibitem{Shuntov:2022qwu}
M.~Shuntov, H.~J.~McCracken, R.~Gavazzi, C.~Laigle, J.~R.~Weaver, I.~Davidzon, O.~Ilbert, O.~B.~Kauffmann, A.~Faisst and Y.~Dubois, \textit{et al.}
``COSMOS2020: Cosmic evolution of the stellar-to-halo mass relation for central and satellite galaxies up to z {\ensuremath{\sim}} 5,''
Astron. Astrophys. \textbf{664}, A61 (2022)
[arXiv:2203.10895 [astro-ph.GA]].

\bibitem{Salucci:2018hqu}
P.~Salucci,
``The distribution of dark matter in galaxies,''
Astron. Astrophys. Rev. \textbf{27}, no.1, 2 (2019)
doi:10.1007/s00159-018-0113-1
[arXiv:1811.08843 [astro-ph.GA]].

\bibitem{Dekel:1986ehj}
A.~Dekel and J.~Silk,
``The origin of dwarf galaxies, cold dark matter, and biased galaxy formation,''
Astrophys. J. \textbf{303}, 39-55 (1986)

\bibitem{Kay:2001hq}
S.~T.~Kay, F.~R.~Pearce, C.~S.~Frenk and A.~Jenkins,
``Including star formation and supernova feedback within cosmological simulations of galaxy formation,''
Mon. Not. Roy. Astron. Soc. \textbf{330}, 113 (2002)
doi:10.1046/j.1365-8711.2002.05070.x
[arXiv:astro-ph/0106462 [astro-ph]].

\bibitem{Efstathiou:1992zz}
G.~Efstathiou,
``Suppressing the formation of dwarf galaxies via photoionization,''
Mon. Not. Roy. Astron. Soc. \textbf{256}, 43P-47P (1992)

\bibitem{Birnboim:2003xa}
Y.~Birnboim and A.~Dekel,
``Virial shocks in galactic haloes?,''
Mon. Not. Roy. Astron. Soc. \textbf{345}, 349-364 (2003)
[arXiv:astro-ph/0302161 [astro-ph]].


\bibitem{Fabian:2012xr}
A.~C.~Fabian,
Ann. Rev. Astron. Astrophys. \textbf{50}, 455-489 (2012)
doi:10.1146/annurev-astro-081811-125521
[arXiv:1204.4114 [astro-ph.CO]].


\bibitem{Croton:2005hbr}
D.~J.~Croton, V.~Springel, S.~D.~M.~White, G.~De Lucia, C.~S.~Frenk, L.~Gao, A.~Jenkins, G.~Kauffmann, J.~F.~Navarro and N.~Yoshida,
``The Many lives of AGN: Cooling flows, black holes and the luminosities and colours of galaxies,''
Mon. Not. Roy. Astron. Soc. \textbf{365}, 11-28 (2006)
[erratum: Mon. Not. Roy. Astron. Soc. \textbf{367}, 864 (2006)]
[arXiv:astro-ph/0602065 [astro-ph]].

\bibitem{Madau:1997pg}
P.~Madau, L.~Pozzetti and M.~Dickinson,
Astrophys. J. \textbf{498} (1998), 106
doi:10.1086/305523
[arXiv:astro-ph/9708220 [astro-ph]].

\bibitem{Kennicutt:1998zb}
R.~C.~Kennicutt, Jr.,
``Star formation in galaxies along the Hubble sequence,''
Ann. Rev. Astron. Astrophys. \textbf{36}, 189-231 (1998)
[arXiv:astro-ph/9807187 [astro-ph]].

\bibitem{Oke:1983nt}
J.~B.~Oke and J.~E.~Gunn,
``Secondary standard stars for absolute spectrophotometry,''
Astrophys. J. \textbf{266}, 713 (1983)

\bibitem{Shapiro:1993hn}
P.~R.~Shapiro, M.~L.~Giroux and A.~Babul,
``Reionization in a cold dark matter universe: The Feedback of galaxy formation on the intergalactic medium,''
Astrophys. J. \textbf{427}, 25 (1994)

\bibitem{Hui:1997dp}
L.~Hui and N.~Y.~Gnedin,
``Equation of state of the photoionized intergalactic medium,''
Mon. Not. Roy. Astron. Soc. \textbf{292}, 27 (1997)
[arXiv:astro-ph/9612232 [astro-ph]].

\bibitem{Barkana:2000fd}
R.~Barkana and A.~Loeb,
``In the beginning: The First sources of light and the reionization of the Universe,''
Phys. Rept. \textbf{349}, 125-238 (2001)
[arXiv:astro-ph/0010468 [astro-ph]].

\bibitem{Springel:2002ux}
V.~Springel and L.~Hernquist,
``The history of star formation in a lcdm universe,''
Mon. Not. Roy. Astron. Soc. \textbf{339}, 312 (2003)
[arXiv:astro-ph/0206395 [astro-ph]].

\bibitem{Okamoto:2008sn}
T.~Okamoto, L.~Gao and T.~Theuns,
``Massloss of galaxies due to a UV-background,''
Mon. Not. Roy. Astron. Soc. \textbf{390}, 920 (2008)
[arXiv:0806.0378 [astro-ph]].

\bibitem{Mesinger:2008ze}
A.~Mesinger and M.~Dijkstra,
``UV Radiative Feedback During the Advanced Stages of Reionization,''
Mon. Not. Roy. Astron. Soc. \textbf{390}, 1071 (2008)
[arXiv:0806.3090 [astro-ph]].


\bibitem{Sobacchi:2014rua}
E.~Sobacchi and A.~Mesinger,
``Inhomogeneous recombinations during cosmic reionization,''
Mon. Not. Roy. Astron. Soc. \textbf{440}, no.2, 1662-1673 (2014)
[arXiv:1402.2298 [astro-ph.CO]].

\bibitem{Sobacchi:2015gpa}
E.~Sobacchi and A.~Mesinger,
``The clustering of Lyman \ensuremath{\alpha} emitters at z \ensuremath{\approx} 7: implications for reionization and host halo masses,''
Mon. Not. Roy. Astron. Soc. \textbf{453}, no.2, 1843-1854 (2015)
[arXiv:1505.02787 [astro-ph.CO]].

\bibitem{Lesgourgues:2011re}
J.~Lesgourgues,
``The Cosmic Linear Anisotropy Solving System (CLASS) I: Overview,''
[arXiv:1104.2932 [astro-ph.IM]].

\bibitem{Blas:2011rf}
D.~Blas, J.~Lesgourgues and T.~Tram,
``The Cosmic Linear Anisotropy Solving System (CLASS) II: Approximation schemes,''
JCAP \textbf{07} (2011), 034
[arXiv:1104.2933 [astro-ph.CO]].

\bibitem{Finkelstein:2022}
S.~L.~Finkelstein and M.~B.~Bagley,
``On the Co-Evolution of the AGN and Star-Forming Galaxy Ultraviolet Luminosity Functions at 3 < z < 9,''
Astrophys. J. \textbf{938}, 25 (2022)
[arXiv:2207.02233 [astro-ph.GA]].


\bibitem{Hastings:1970aa}
W.~K.~Hastings,
``Monte Carlo Sampling Methods Using Markov Chains and Their Applications,''
Biometrika \textbf{57} (1970), 97-109

\bibitem{Audren:2012wb}
B.~Audren, J.~Lesgourgues, K.~Benabed and S.~Prunet,
``Conservative Constraints on Early Cosmology: an illustration of the Monte Python cosmological parameter inference code,''
JCAP \textbf{02} (2013), 001
[arXiv:1210.7183 [astro-ph.CO]].

\bibitem{Brinckmann:2018cvx}
T.~Brinckmann and J.~Lesgourgues,
``MontePython 3: boosted MCMC sampler and other features,''
Phys. Dark Univ. \textbf{24} (2019), 100260
[arXiv:1804.07261 [astro-ph.CO]].


\bibitem{Planck:2018vyg}
N.~Aghanim \textit{et al.} [Planck],
``Planck 2018 results. VI. Cosmological parameters,''
Astron. Astrophys. \textbf{641}, A6 (2020)
[erratum: Astron. Astrophys. \textbf{652}, C4 (2021)]
[arXiv:1807.06209 [astro-ph.CO]].

\bibitem{Pan-STARRS1:2017jku}
D.~M.~Scolnic \textit{et al.} [Pan-STARRS1],
``The Complete Light-curve Sample of Spectroscopically Confirmed SNe Ia from Pan-STARRS1 and Cosmological Constraints from the Combined Pantheon Sample,''
Astrophys. J. \textbf{859}, no.2, 101 (2018)
[arXiv:1710.00845 [astro-ph.CO]].

\bibitem{supp_mat}
See Supplemental Material at [URL will be inserted by publisher] for numerical values of the constraints derived here.

\bibitem{Williams:2024}
C.~E.~Williams, W.~Lake, S.~Naoz, B.~Burkhart, T.~Treu, F.~Marinacci, Y.~Nakazato, M.~Vogelsberger, N.~Yoshida, G.~Chiaki, Y.~S.~Chiou and A.~Chen,
``The Supersonic Project: Lighting Up the Faint End of the JWST UV Luminosity Function,''
Astrophys. J. Lett. \textbf{960}, no.2, L16 (2024)
[arXiv:2310.03799 [astro-ph.GA]].

\bibitem{Rasmussen:2006}
C.~E.~Rasmussen and C.~K.~I.~Williams,
``Gaussian Processes for Machine Learning'',
MIT Press, Cambridge, USA (2006),
ISBN: 978-0-262-18253-9. 


\bibitem{Stark:2025aaa}
Stark, D.~P., Topping, M.~W., Endsley, R., Tang, M.
``Observations of the First Galaxies in the Era of JWST,'' 
[arXiv:2501.17078 [astro-ph.CO]].


\bibitem{Donnan:2024}
C.~T.~Donnan, R.~J.~McLure, J.~S.~Dunlop, D.~J.~McLeod, D.~Magee, K.~Z.~Arellano-C{\'o}rdova, L.~Barrufet, R.~Begley, R.~A.~A.~Bowler, A.~C.~Carnall, F.~Cullen, R.~S.~Ellis, A.~Fontana, G.~D.~Illingworth, N.~A.~Grogin, M.~L.~Hamadouche, A.~M.~Koekemoer, F.~Y.~Liu, C.~Mason, P.~Santini and T.~M.~Stanton,
``JWST PRIMER: a new multifield determination of the evolving galaxy UV luminosity function at redshifts z $\simeq$ 9-15,''
Mon. Not. Roy. Astron. Soc. \textbf{533}, no.3, 3222--3237 (2024)
[arXiv:2403.03171 [astro-ph.GA]].


\bibitem{Harikane:2023}
Y.~Harikane, M.~Ouchi, M.~Oguri, Y.~Ono, K.~Nakajima, Y.~Isobe, H.~Umeda, K.~Mawatari and Y.~Zhang,
``A Comprehensive Study of Galaxies at z $\simeq$ 9-16 Found in the Early JWST Data: Ultraviolet Luminosity Functions and Cosmic Star Formation History at the Pre-reionization Epoch,''
Astrophys. J. Suppl. \textbf{265}, no.1, 5 (2023)
[arXiv:2208.01612 [astro-ph.GA]].
m

\bibitem{Yung:2023bng}
L.~Y.~A.~Yung, R.~S.~Somerville, S.~L.~Finkelstein, S.~M.~Wilkins and J.~P.~Gardner,
``Are the ultra-high-redshift galaxies at z {\ensuremath{>}} 10 surprising in the context of standard galaxy formation models?,''
Mon. Not. Roy. Astron. Soc. \textbf{527} (2023) no.3, 5929-5948
[arXiv:2304.04348 [astro-ph.GA]].

\bibitem{DeBoer:2016tnn}
D.~R.~DeBoer, A.~R.~Parsons, J.~E.~Aguirre, P.~Alexander, Z.~S.~Ali, A.~P.~Beardsley, G.~Bernardi, J.~D.~Bowman, R.~F.~Bradley and C.~L.~Carilli, \textit{et al.}
``Hydrogen Epoch of Reionization Array (HERA),''
Publ. Astron. Soc. Pac. \textbf{129}, no.974, 045001 (2017)
[arXiv:1606.07473 [astro-ph.IM]].



\bibitem{LOFAR:2013jil}
M.~P.~van Haarlem \textit{et al.} [LOFAR],
``LOFAR: The LOw-Frequency ARray,''
Astron. Astrophys. \textbf{556}, A2 (2013)
[arXiv:1305.3550 [astro-ph.IM]].

\bibitem{Mellema:2012ht}
G.~Mellema, L.~V.~E.~Koopmans, F.~A.~Abdalla, G.~Bernardi, B.~Ciardi, S.~Daiboo, A.~Ferrara, I.~T.~Iliev, V.~Jelić, M.~J.~Jarvis, P.~Lazio, A.~R.~Offringa, V.~Pandey, J.~Schaye, R.~M.~Thomas, S.~J.~Wijnholds, H.~J.~A.~Röttgering and S.~Zaroubi,
``Reionization and the Cosmic Dawn with the Square Kilometre Array,''
Exp. Astron. \textbf{36}, 235–318 (2013)
[arXiv:1210.0197 [astro-ph.CO]].



\bibitem{Flitter:2023mjj}
J.~Flitter and E.~D.~Kovetz,
``New tool for 21-cm cosmology. I. Probing \ensuremath{\Lambda}CDM and beyond,''
Phys. Rev. D \textbf{109} (2024) no.4, 043512
[arXiv:2309.03942 [astro-ph.CO]].

\bibitem{Rahimieh:2025fsb}
A.~Rahimieh, P.~Parashari, R.~An, T.~Driskell, J.~Mirocha and V.~Gluscevic,
``Sensitivity of the Global 21-cm Signal to Dark Matter-Baryon Scattering,''
[arXiv:2505.03148 [astro-ph.CO]].


\bibitem{Rahimieh:2025lbf}
A.~Rahimieh, P.~Parashari and V.~Gluscevic,
``Forecasting 21-cm power spectrum sensitivity to dark Matter-baryon scattering,''
[arXiv:2508.20507 [astro-ph.CO]].

\bibitem{Munoz:2019hjh}
J.~B.~Mu{\~n}oz, C.~Dvorkin and F.~Y.~Cyr-Racine,
``Probing the Small-Scale Matter Power Spectrum with Large-Scale 21-cm Data,''
Phys. Rev. D \textbf{101} (2020) no.6, 063526
[arXiv:1911.11144 [astro-ph.CO]].





























































\end{thebibliography}
\end{document}